\documentclass[a4paper]{article}
\let\oldtabular\tabular
\renewcommand{\tabular}{\footnotesize\oldtabular}
\usepackage[english]{babel}
\usepackage[utf8]{inputenc}
\usepackage{amsmath}
\usepackage{graphicx}
\usepackage[colorinlistoftodos]{todonotes}
\usepackage{algorithm}
\usepackage{algpseudocode}
\usepackage{amssymb}
\usepackage{nomencl}
\usepackage{etoolbox}
\renewcommand\nomgroup[1]{%
  \item[\bfseries
  \ifstrequal{#1}{R}{Unit Commitment}{%
  \ifstrequal{#1}{D}{Data-Driven Approach}{%
  \ifstrequal{#1}{A}{Acronyms}{%
  \ifstrequal{#1}{F}{Frequency Dynamics}{}}}}%
]}
\makenomenclature
\usepackage{eurosym}
\usepackage{indentfirst}
\usepackage{nicefrac}
\usepackage{booktabs}
\usepackage{url}
\usepackage[hidelinks]{hyperref}
\usepackage{cleveref} 
\usepackage{mathtools} 

\title{Inclusion of Frequency Nadir constraint in the Unit Commitment Problem of Small Power Systems Using Machine Learning}

\author{Mohammad Rajabdorri, Behzad Kazemtabrizi, Matthias Troffaes, Lukas Sigrist, Enrique Lobato}

\date{\today}

\begin{document}
\maketitle

\begin{abstract}
As the intention is to reduce the amount of thermal generation and to increase the share of clean energy, power systems are increasingly becoming susceptible to frequency instability after outages due to reduced levels of inertia. To address this issue frequency constraints are being included in the scheduling process, which ensure a tolerable frequency deviation in case of any contingencies. In this paper, a method is proposed to integrate the non-linear frequency nadir constraint into the unit commitment problem, using machine learning. First a synthetic training dataset is generated. Then two of the available classic machine learning methods, namely logistic regression and support vector machine, are proposed to predict the frequency nadir. To be able to compare the machine learning methods to traditional frequency constrained unit commitment approaches, simulations on the power system of La Palma island are carried out for both proposed methods as well as an analytical linearized formulation of the frequency nadir. Our results show that the unit commitment problem with a machine learning based frequency nadir constraint is solved considerably faster than with the analytical formulation, while still achieving an acceptable frequency response quality after outages.
\end{abstract}
\nomenclature[R]{\(suc(.)\)}{start-up costs [\euro]}
\nomenclature[R]{\(gc\)}{generation costs [\euro]}
\nomenclature[R]{\(u\)}{commitment variable [$\in$\{0,1\}]}
\nomenclature[R]{\(v\)}{start-up variable [$\in$\{0,1\}]}
\nomenclature[R]{\(w\)}{shut-down variable [$\in$\{0,1\}]}
\nomenclature[R]{\(p\)}{power generation variable [MW]}
\nomenclature[R]{\(t\)}{index of time intervals}
\nomenclature[R]{\(T\)}{set of all time intervals}
\nomenclature[R]{\(i\)}{index of generators}
\nomenclature[R]{\(\mathcal{I}\)}{set of all generators}
\nomenclature[R]{\(s\)}{alias index for time intervals}
\nomenclature[R]{\(ii\)}{alias index for generators}
\nomenclature[R]{\(UT\)}{minimum up-time of generators [hours]}
\nomenclature[R]{\(DT\)}{minimum down-time of generators [hours]}
\nomenclature[R]{\(\overline{\mathcal{P}_i}\)}{maximum power output of generator $i$ [MW]}
\nomenclature[R]{\(\underline{\mathcal{P}_i}\)}{minimum power output of generator $i$ [MW]}
\nomenclature[R]{\(\overline{\mathcal{R}_i}\)}{maximum ramp-up of generator $i$ [MW/h]}
\nomenclature[R]{\(\underline{\mathcal{R}_i}\)}{maximum ramp-down of generator $i$ [MW/h]}
\nomenclature[R]{\(r\)}{online reserve power variable [MW]}
\nomenclature[R]{\(wg\)}{wind generation variable [MW]}
\nomenclature[R]{\(sg\)}{solar generation variable [MW]}
\nomenclature[R]{\(I\)}{number of generators}
\nomenclature[R]{\(\overline{\mathcal{D}}\)}{maximum yearly thermal generation}
\nomenclature[R]{\(\underline{\mathcal{D}}\)}{minimum yearly thermal generation}
\nomenclature[R]{\(\underline{\mathcal{D}_t}\)}{demand at hour $t$}
\nomenclature[D]{\(K_i\)}{number of the steps}
\nomenclature[D]{\(FC\)}{set of feasible combinations}
\nomenclature[D]{\(x\)}{features of the dataset}
\nomenclature[D]{\(y\)}{labels of the dataset}
\nomenclature[D]{\(\mathcal{X}\)}{set of all features}
\nomenclature[D]{\(\mathcal{Y}\)}{set of all labels}
\nomenclature[D]{\(f_\theta(x)\)}{hypothesis function}
\nomenclature[D]{\(\theta\)}{coefficients in the linear model}
\nomenclature[D]{\(\Theta\)}{set of $\theta$ parameters}
\nomenclature[D]{\(\ell(.)\)}{loss function}
\nomenclature[D]{\(\hat{y}\)}{predicted label}
\nomenclature[D]{\(N\)}{number of data samples}
\nomenclature[D]{\(n\)}{index of data samples}
\nomenclature[D]{\(M\)}{number of features}
\nomenclature[D]{\(m\)}{index of features}
\nomenclature[D]{\(C\)}{regularization coefficient}
\nomenclature[F]{\(H\)}{inertia [MW.s]}
\nomenclature[F]{\(f_0\)}{nominal frequency [Hz]}
\nomenclature[F]{\(f(t)\)}{frequency [Hz]}
\nomenclature[F]{\(D\)}{load damping factor [\%/Hz]}
\nomenclature[F]{\(P_m\)}{mechanical power [MW]}
\nomenclature[F]{\(P_e\)}{electrical power [MW]}
\nomenclature[F]{\(\mathcal{M}\)}{base power of the unit [MW]}
\nomenclature[F]{\(\ell\)}{index of the lost generator}
\nomenclature[F]{\(P_\ell\)}{lost power [MW]}
\nomenclature[F]{\(\Delta f_{crit}^{'}\)}{critical rate of change of frequency}
\nomenclature[F]{\(\Delta f_{crit}^{ss}\)}{critical steady state frequency [Hz]}
\nomenclature[F]{\(T_g\)}{delivery time of units [s]}
\nomenclature[F]{\(\Delta f_{crit}^{nadir}\)}{critical frequency nadir [Hz]}
\nomenclature[F]{\(a_j\)}{breaking point}
\nomenclature[F]{\(j\)}{breaking point index}
\nomenclature[F]{\(J\)}{number of the breaking points}
\nomenclature[F]{\(\lambda_j\)}{weight associated with breaking point $j$}
\nomenclature[F]{\(\gamma_j\)}{binary operator of affine segments [$\in$\{0,1\}]}
\nomenclature[F]{\(\alpha, \beta\)}{normalizing coefficients}
\nomenclature[F]{\(z_1, z_2\)}{auxiliaries for changing variables}

\printnomenclature

\section{Introduction}\label{introduction}

The share of renewable energy sources (RES) is growing steadily in power systems. It is essential to facilitate the growth of RES penetration to reduce the carbon emission from fossil fuel based generators. There are however some obstacles that limit the applicability of RES. RES are volatile in nature and forecasting them is subject to uncertainties. Hence, integrating them in the power system is challenging. Moreover RES are usually decoupled from the system, and therefore they do not add any inertia to the system. This is particularly important in small power systems like islands, as they typically suffer from inertia scarcity, and are therefore more prone to frequency volatility. For that reason, when integrating RES in such systems, it can be very challenging to maintain frequency stability in case of contingencies.

To address this issue, researchers have included frequency dynamics in short term scheduling processes like Unit Commitment (UC) to form a frequency constrained UC (FCUC), in \cite{trovato2018unit}, \cite{badesa2019simultaneous}, \cite{paturet2020stochastic}, and etc. The standard (non-frequency constrained) UC problem can be formulated as a mixed integer linear programming (MILP) problem, which can be solved efficiently using standard solvers. Unfortunately, the frequency dynamics of a power system is highly nonlinear and non-convex, complicating how the UC problem can still be formulated as a MILP problem. There is valuable research work in the literature, addressing this very issue (\cite{mousavi2020integration}, \cite{rabbanifar2020frequency}, \cite{shahidehpour2021two}, and \cite{Ferrandon-Cervantes2022}). Frequency dynamics after outages are usually described by the rate of change of frequency (RoCoF), frequency nadir, and steady-state frequency. RoCoF and steady-state frequency can be formulated linearly, but frequency nadir cannot. In previously mentioned studies, the non-linear constraint on the frequency nadir (derived from the well-known swing equation) has been simplified or approximated so that it still can be used in the MILP formulation of UC problem. These formulations are based on simplifying assumptions and usually are  computationally demanding. More recently, data-driven approaches are being introduced to more accurately model the frequency dynamics in the UC problem, instead of relying on analytical methods (\cite{lagos2021data}, \cite{Zhang2021}, \cite{Zhang2021a}, \cite{Yang2022}). These methods try to estimate the dynamics of the system accurately, while keeping the solution time of UC reasonably low.

Among the analytical methods, in \cite{trovato2018unit}, a linear formulation of inertial response and the frequency response of the system is added to the UC problem, which makes sure that in case of the largest outage, there is enough ancillary service to prevent under frequency load shedding (UFLS). To linearize frequency nadir constraint, first-order partial derivatives of its equation with respect to higher-order non-linear variables are calculated. Then the frequency nadir is presented by a set of piecewise linear constraints. In \cite{badesa2019simultaneous}, different frequency services are optimized simultaneously with a stochastic unit commitment (SUC) approach, targeting low inertia systems that have high levels of RES penetration. The stochastic model uses scenario trees, generated by the quantile-based scenario generation method. To linearize frequency nadir, an inner approximation method is used for one side of the constraint, and for the other side, a binary expansion is employed to approximate the constraint as a MILP using the big-M technique.
In \cite{paturet2020stochastic}, a stochastic unit commitment approach is introduced for low inertia systems, that includes frequency-related constraints. The problem considers both the probability of failure events and wind power uncertainty to compute scenario trees for the two-stage SUC problem. An alternative linearization approach is used to make sure the nadir threshold is not violated. Instead of piece-wise linearizing the whole equation, relevant variables including the nonlinear equation are confined within a plausible range that guarantees frequency drop after any contingency will be acceptable. In \cite{mousavi2020integration}, a reformulation linearization technique is employed to linearize the frequency nadir limit equation. Results show that controlling the dynamic frequency during the scheduling process decreases the operation costs of the system while ensuring its frequency stability. In \cite{rabbanifar2020frequency}, first, a frequency response model is developed that provides enough primary frequency response and system inertia in case of any outages. All frequency dynamic metrics, including the RoCoF and frequency nadir are obtained from this model, as analytic explicit functions of UC state variables and generation loss. These functions are then linearized based on a pseudo-Boolean theorem, so they can be implemented in linear frequency constrained UC problem. To find the optimal thermal unit commitment and virtual inertia placement, a two-stage chance-constrained stochastic optimization method is introduced in \cite{shahidehpour2021two}. Frequency nadir is first calculated with a quadratic equation and then it is constrained with the help of the big-M approach. In \cite{Ferrandon-Cervantes2022}, the frequency nadir is approximated as a piece-wise linear function to good (and in principle, arbitrary) precision, and the resulting constraint is then reformulated as a MILP using separable programming. A common assumption in \cite{trovato2018unit}, \cite{badesa2019simultaneous}, \cite{paturet2020stochastic}, \cite{shahidehpour2021two}, \cite{Ferrandon-Cervantes2022}, and many other similar works, is that the provision of reserve increases linearly in time, and all units will deliver their available reserve within a given fixed time. This assumption is the key to calculate the frequency nadir as a function of other variables.

\begin{figure}[!htbp]
    \centering
    \includegraphics[width=0.7\linewidth]{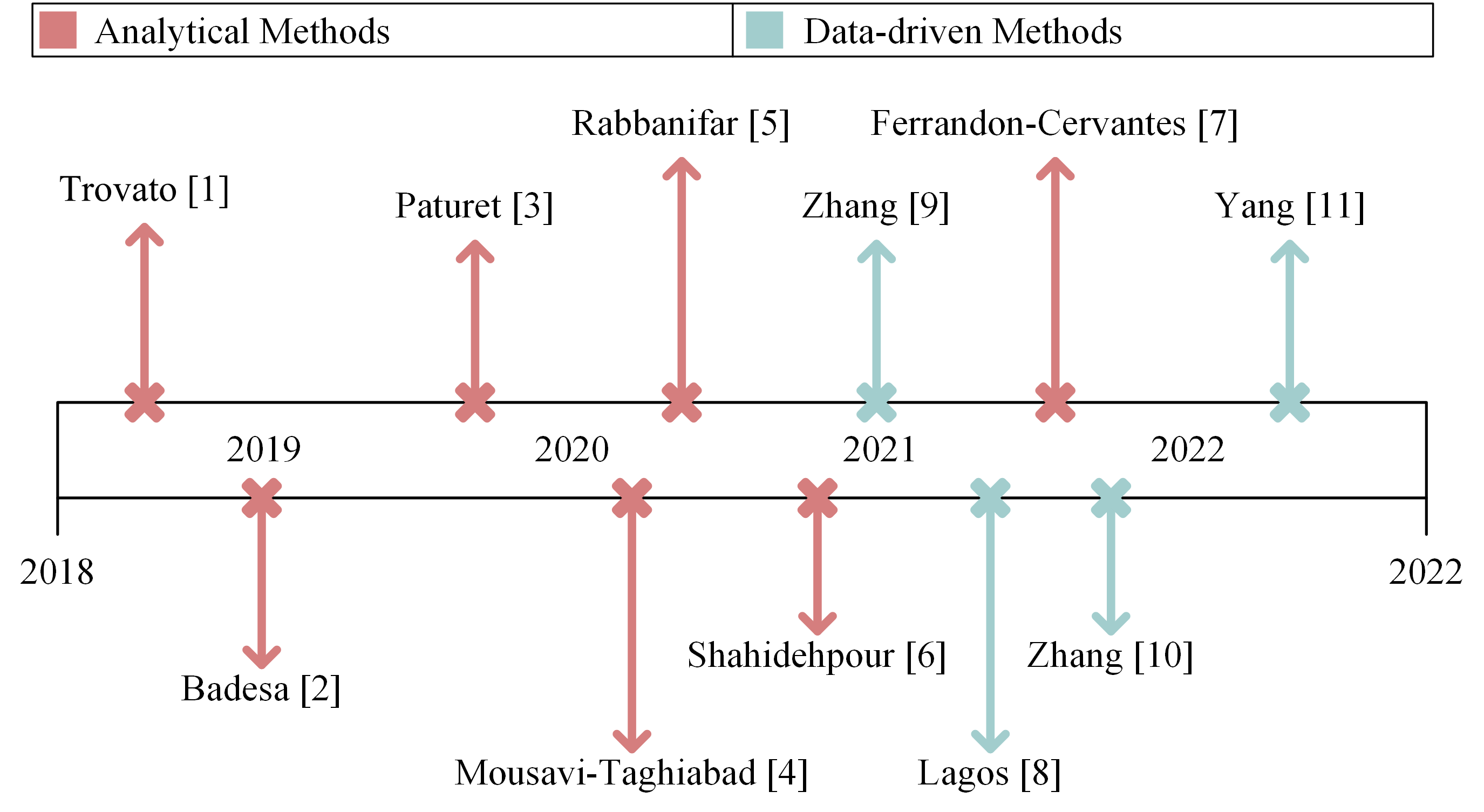}
    \caption{Summary of the reviewed literature}
    \label{fig:refrences}
\end{figure}

Among the data-driven approaches, in \cite{lagos2021data} a multivariate optimal classification trees (OCT) technique is used to learn linear frequency constraints. A robust formulation is proposed to address the uncertainties of load and RES. OCT is defined and solved as an MILP optimization problem, so if the training dataset is big, optimizing the OCT becomes very hard. A dynamic model is presented in \cite{Zhang2021} to generate the training dataset. The generated dataset is trained, using a deep neural network. Trained neural networks are formulated so they can be used in an MIL problem and the frequency nadir predictor is developed, to be used in UC problem. Then in \cite{Zhang2021a} deep neural network (DNN) is trained by high-fidelity power simulation and reformulated as an MIL set of constraints to be used in UC. The generated data samples in \cite{Zhang2021a} are denser where the frequency nadirs are closer to the UFLS threshold. In \cite{Yang2022} linear regression is applied on a synthetic training dataset to extract the relationship between frequency response and frequency deviation during primary frequency response. The obtained regression is then used as a constraint in a distributionally robust economic dispatch model. The results of these data-driven methods is heavily reliant on the quality of the training dataset. Also, defining the DNN as MIL constraints to the UC problem, adds so many variables and sets of constraints to the formulation. A summary of the reviewed papers is presented in \cref{fig:refrences}.

Following the same line of research, this paper generates a synthetic training dataset, and then applies machine learning (ML) methods on the dataset to derive a linear constraint which approximates the original non-linear frequency nadir constraint for all scenarios in the dataset. To evaluate the effectiveness of the ML methods, weekly FCUC of La Palma island power system is solved for seasonal sample weeks. The results are compared to one of the recent FCUC formulations that employs a MILP formulation based on an analytical expression of the frequency nadir \cite{Ferrandon-Cervantes2022}.  The contributions and highlights of this paper are as follows,
\begin{itemize}
    \item A novel synthetic data generation algorithm is presented that includes feasible operating points. These operating points are sorted by their quadratic generation cost function. Operating points that are cost-efficient are added to the training dataset. Such dataset is only composed of operating points that are close to the optimal solution of the UC problem, as they are feasible and cheaper. Comparing to the proposed datasets in the literature, the data generation algorithm here covers all the feasible generator operating points that are likely to be picked by the solver. An advantage of this dataset, is that it's not sensitive to the daily demand and RES forecast. So once it's generated, the models that are obtained from it, can be used throughout the year.
    \item The training dataset is passed through different ML methods to train a linear constraint that can classify tolerable and intolerable frequency nadirs after any outage. This linear constraint can be directly employed in the UC problem as the frequency nadir constraint.
    \item The performance of the ML methods is compared to the analytical method of predicting frequency nadir, showing that the UC is solved considerably faster, whilst still achieving an acceptable frequency response quality after outages ($N-1$ security criteria). It should be noted that reducing computation times is critical when modelling uncertainties in stochastic or robust UC problems.
\end{itemize}
The rest of this paper is organised as follows: the underlying methodologies used in this paper are presented in \cref{methodology}. Meanwhile, the simulation results and corresponding comparisons are discussed in \cref{results}. Finally the conclusions are drawn in \cref{conclusion}.

\section{Methodology}\label{methodology}

In this section a general UC formulation is presented. To incorporate the frequency dynamics in the UC problem, rate of change of frequency (in \cref{sec:rocof}), steady-state frequency (in \cref{sec:qss}), and frequency nadir (in \cref{sec:fNadir}) constraints are formulated. Frequency nadir is incorporated with both analytical (in \cref{sec:analytical}) and ML based (in \cref{sec:ML}) methods. Finally, in section 2.6, the two options (analytic and ML modelling of frequency nadir) are compared in practice using the FCUC results. 
\subsection{UC Formulation}\label{sec:UCF}

The Unit Commitment (UC) problem is a mixed-integer problem and is usually solved with MILP solvers after linearization of non-linear terms, introducing additional integer or binary auxiliary variables where need be to handle non-linearities. A general representation of the UC problem is provided here,
\begin{subequations} \label{e7}
\begin{align}
\min_{x,p}&\ suc(u_{t,i})+gc(p_{t,i})\tag{\ref{e7}}
\\
u_{t,i}-u_{t-1,i}&=v_{t,i}-w_{t,i} &&\text{\scriptsize $t\in T,\;i\in\mathcal{I}$}\label{e8a}
\\
v_{t,i}+w_{t,i}&\leq1 &&\text{\scriptsize $t\in T,\;i\in\mathcal{I}$}\label{e8b}
\\
\sum_{s=t-UT_g+1}^{t}v_{s,i}&\leq u_{t,i} &&\text{\scriptsize $t\in\{UT_g,\dots, T\}$}\label{e8c}
\\
\sum_{s=t-DT_g+1}^{t}w_{s,i}&\leq 1-u_{t,i}&& \text{\scriptsize $t\in\{UT_g,\dots, T\}$}\label{e8d}
\\
p_{t,i}&\geq \underline{\mathcal{P}}_i u_{t,i} &&\text{\scriptsize $t\in T,\;i\in\mathcal{I}$}\label{e9a}
\\
p_{t,i}+r_{t,i}&\leq \overline{\mathcal{P}}_i u_{t,i} &&\text{\scriptsize $t\in T,\;i\in\mathcal{I}$}\label{e9b}
\\
p_{t-1,i}-p_{t,i}&\leq \underline{\mathcal{R}}_i &&\text{\scriptsize $t\in T,\;i\in\mathcal{I}$}\label{e9c}
\\
p_{t,i}-p_{t-1,i}&\leq \overline{\mathcal{R}}_i &&\text{\scriptsize $t\in T,\;i\in\mathcal{I}$}\label{e9d}
\\
\sum_{i\in\mathcal{I}}\big(p_{t,i}\big)+wg_t+sg_t&=\mathcal{D}_t &&\text{\scriptsize $t\in T$}\label{e9e}
\end{align}
\end{subequations}
The aim is to solve \cref{e7} subject to \cref{e8a,e8b,e8c,e8d,e9a,e9b,e9c,e9d,e9e}. $gc(\cdot)$ is usually a convex cost function, which can be easily piece-wise linearized to turn it into a MILP problem.  \Cref{e8a,e8b} represent the binary logic of the UC problem. \Cref{e8c,e8d} are the minimum up-time and minimum downtime constraints of the units. \Cref{e9a} is the minimum power generation constraint. \Cref{e9b} is the maximum power generation constraint, and states that the summation of power generation and power reserve of every online unit, should be less than the maximum output of the unit. \Cref{e9c,e9d} are ramp-down and ramp-up constraints. \Cref{e9e} is the power balance equation.

\subsection{Frequency Dynamics}

The dynamics of the generator rotor is usually described by the swing equation,
\begin{equation}\label{eq:swing}
\frac{2H}{f_0}\frac{d\Delta f(t)}{dt}+D\mathcal{D}_t\Delta f(t)=P_m-P_e
\end{equation}
This is a first order differential equation. When an outage happens, there will be a power mismatch between the mechanical output of the units and the electrical demand, which is equal to the amount of lost power.
\begin{equation}\label{eq:mismatch}
P_\ell\coloneqq P_m-Pe
\end{equation}
The available inertia after the outage of unit $\ell$ can be defined as,
\begin{equation}\label{eq:inertia}
\mathcal{H}_\ell\coloneqq\sum_{i\in \mathcal{I}, i\neq \ell}(H_i\mathcal{M}_i u_{t,i})
\end{equation}
Considering the swing equation and the operating point of the system before the outage, the frequency response of the system after the outage can be calculated. The frequency response is reflected in metrics like Rate of Change of Frequency (RoCoF), steady state frequency, and frequency nadir.

\subsection{Rate of Change of Frequency Modelling}\label{sec:rocof}

The RoCoF after the outage can be drived from \cref{eq:swing}. The amount of inertia after the outage should be able to prevent exceeding from critical RoCoF,
\begin{equation}\label{eq:rocof}
\mathcal{H}_\ell\geq \frac{P_\ell f_0}{2\Delta f_{crit}^{'}}\;\;\text{\scriptsize $t\in T,\;\forall \ell$}
\end{equation}
This equation is linear and can be directly added to the MILP for the UC problem.

\subsection{Steady State Frequency Modelling}\label{sec:qss}

For the steady state frequency after an outage, it is assumed that frequency is converged and there has been enough time for units to deliver their reserve power. To make sure that the steady state frequency is not violated, this constraint can be derived from the swing equation,
\begin{equation}\label{eq:ssf}
\sum\limits_{i \in \mathcal{I}, i\neq \ell}r_{t,i}\geq P_\ell-D\mathcal{D}_t\Delta f^{ss}_{crit}\;\;\text{\scriptsize $t\in T,\;\forall \ell$}
\end{equation}
This is also linear and can be directly added to the MILP.

\subsection{Frequency Nadir Modelling}\label{sec:fNadir}

The inclusion of frequency nadir into the MILP is more complicated, as it is non-convex. Traditionally researchers have proposed analytical methods to calculate frequency nadir from the swing equation, and then implement a MILP approximation of it to the UC problem, which is discussed in \cref{sec:analytical}. More recently the use of ML is gaining popularity to model these more complicated situations. The proposed method of this study to estimate frequency nadir is presented in \cref{sec:ML}.

\subsubsection{Frequency nadir: analytical modelling}\label{sec:analytical}

After the outage happens the frequency starts to decrease. As a response the remaining units start ramping up, if they have any available reserve. Here, and in \cite{trovato2018unit,badesa2019simultaneous,paturet2020stochastic,shahidehpour2021two,Ferrandon-Cervantes2022} it is assumed that the reserve power of each unit is delivered linearly and will reach to its maximum output power in $T_g$ seconds. This is a key assumption to finding an analytical expression for the frequency nadir.
\begin{equation}\label{eq:reslin}
r_{t,i}(\tau) =
    \begin{cases}
      \frac{r_{t,i}\tau}{T_g} & \text{if $t\leq T_g$}\\
      r_{t,i} & \text{if $t > T_g$}
    \end{cases}       
\end{equation}
It is also assumed that frequency nadir happens before $T_g$. Let's also define the amount of remaining reserve after the outage of unit $\ell$ as,
\begin{equation}\label{eq:reserveAO}
    \mathcal{R}_\ell=\sum_{i \in \mathcal{I}, i\neq \ell}r_{t,i}
\end{equation}
With these assumptions the frequency nadir constraint is as follows,
\begin{equation}\label{eq:nadir}
    \mathcal{H}_\ell\mathcal{R}_\ell - \frac{f_0T_gP_\ell^2}{4\Delta f^{nadir}_{crit}}+\frac{D\mathcal{D}_tT_gP_\ell f_0}{4}\geq0\;\;\text{\scriptsize $t\in T,\;\forall \ell$}
\end{equation}
This constraint cannot be added to the MILP formulation of UC, because it is non-convex, due to the product of inertia and reserve. In \cite{Ferrandon-Cervantes2022}, auxiliary variables and separable programming are introduced to linearize these terms. The following constraints are used to linearize the square of lost power,
\begin{subequations}\label{eq:linearization}
\begin{align}
        P_\ell&=\sum_{j=0}^{J}a_j\lambda_j
    \\
        P_\ell^2&\approx \sum_{j=0}^{J}(a_j)^2\lambda_j
    \\
        \sum_{j=0}^{J}\lambda_j&=1
    \\
        \sum_{j=1}^{J}\gamma_j&=1
    \\
        \lambda_0&\leq\gamma_1
    \\
    \lambda_j &\leq \gamma_j +\gamma_{j+1} &&j\in \{1,...,J-1\}
    \\
    \lambda_J&\leq \gamma_J
\end{align}
\end{subequations}
Here, the $a_j$ are fixed constants that control the approximation.
To linearize the production of inertia and reserve with the same manner, first a change in variables should be applied,
\begin{subequations}
\begin{equation}
    \mathcal{H}_i\alpha\mathcal{R}_\ell\beta=z_1^2-z_2^2
\end{equation}
\begin{equation}
    \frac{z_1+z_2}{\alpha}=\mathcal{H}_i
\end{equation}
\begin{equation}
    \frac{z_1-z_2}{\beta}=\mathcal{R}_\ell
\end{equation}
\end{subequations}
Now the new variables $z_1$ and $z_2$ can be used instead of inertia and reserve and their square form can be linearized similar to $P_\ell^2$.

\subsubsection{Frequency Nadir: ML based modelling}\label{sec:ML}

ML methods entail different components: (I) Data, which is a collection of data points that are characterized by features, (II) Model, which consists of feasible hypothesis maps from feature space to label space, (III) Loss function to measure the quality of the model, (IV) and a process of model validation to asses its performance. Each of these topics are discussed in the following.

\paragraph{Data Generation:}

A proper set of data is needed from which to learn the frequency nadir. The training dataset comprises of features $x \in \mathcal{X}$ and labels $y \in \mathcal{Y}$. In case of implementing frequency nadir in the UC problem, features are extracted from operating points and labels are obtained from the frequency nadir measurements after outages. These measurements can be obtained by solving high order differential swing equation, or by using dynamic system frequency (SFR) models. Assigned labels can be numeric (for example the frequency nadir measurement in Hz) or categorical (for example a binary label of whether the obtained frequency nadir is tolerable or not). The features should be chosen wisely so they represent a reasonable amount of information about their labels. On the other hand, unnecessarily large number of features can be detrimental in both computational and statistical aspects. Computationally, choosing a large feature vector increases the dimensions of the problem, so more resources are needed for the calculations. Statistically, using higher number of features makes the model more susceptible to overfitting. It is beneficial to only use features with the most relevant information to predict the label $y$ \cite{Jung2018}. In this paper $y$ is binary, so the proposed ML methods are binary classifications. In the literature different methods are introduced to reduce the size of feature vector. For the purpose of this paper, features must be accessible throughout the UC optimization process. Therefore, the variables that are correlated most with the label will be picked as the features. As will be shown later in \cref{results}, the selected features for predicting frequency nadir adequacy are available inertia after outage ($\mathcal{H}_\ell$), weighted gain of turbine-governor model ($\mathcal{K}_\ell$), the amount of lost generation ($P_\ell$), and the amount of available reserve ($\mathcal{R}_\ell$).

\begin{figure}[!htbp]
\centering
\includegraphics[width=0.7\linewidth]{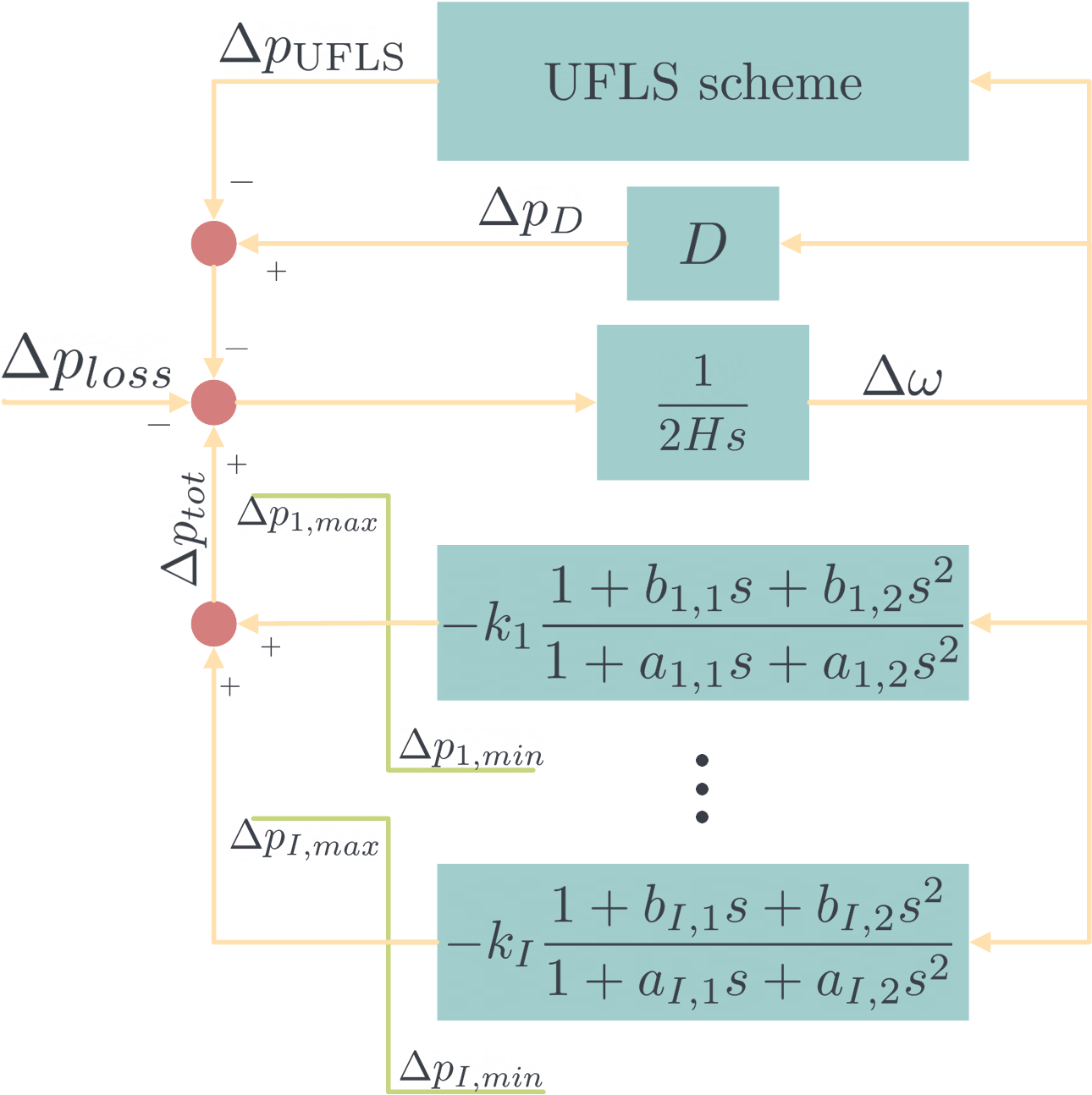}
\caption{SFR model.}
\label{fSFRmodel}\end{figure}

To have a complete dataset, one approach is to consider every combination of possible generation outputs of the units. But many of these combinations are infeasible as they do not satisfy all UC constraints (power balance, reserve constraint, or maximum RoCoF), or unappealing as the optimization problem will favor cheaper combinations. In this paper a data generation method is used, to only generate feasible control points that are cost effective, hence more probable to be close to the solution of the UC optimization problem. The process is listed in \cref{algori}.
\begin{algorithm}[!htbp]
\caption{Synthetic Data Generation}
\label{algori}
\textbf{Input}: for each generator $i\in\{1,\dots,I\}$, a vector of power levels $(p_i^1,p_i^2,\dots,p_i^{K_i})$ where $p_i^1=\underline{\mathcal{P}}_i$ and $p_i^{K_i}=\overline{\mathcal{P}}_i$, \\
lower and upper bounds for total generation: $\underline{\mathcal{D}}$, $\overline{\mathcal{D}}$\\
\textbf{Output}: All feasible and cheap combinations
\begin{algorithmic}[1]
    \For{$(k_1,\dots,k_I)\in \bigtimes_{i=1}^I \{0,\dots,K_i\}$} \Comment{\texttt{\small Combinations of power levels}}
    \For{$i\in\{1,\dots,I\}$}
    \State $u_i\coloneqq 0\text{ if }k_i=0\text{ else }1$ \Comment{\texttt{\small Unit $i$ status}}
    \EndFor
    \State $R\coloneqq \sum_{i=1}^I (p_i^{K_i}u_i-p_i^{k_i})$ \Comment{\texttt{\small Total reserve}}
    \State $G\coloneqq\sum_{i=1}^I p_i^{k_i}$ \Comment{\texttt{\small Total generation}}
    \State $H\coloneqq\sum_{i=1}^I(H_i\mathcal{M}_i u_{i})$\Comment{\texttt{\small Total inertia}}
    \If{$G\in [\underline{\mathcal{D}},\overline{\mathcal{D}}]$ and $R\ge\max_{i=1}^I p_i^{K_i}$ and $H\geq \frac{p_i^{k_i} f_0}{2\Delta f_{crit}^{'}}$}
    \State $FC\gets FC\cup (p_1^{k_1},\dots,p_I^{k_I})$ \Comment{\texttt{\small Combination is feasible}}
    \EndIf
    \EndFor
    \State Sort $FC$ by the quadratic generation cost function
    \State Keep a reasonable number of cheaper combinations and remove the rest
\end{algorithmic}
\end{algorithm}
First vectors of power output of each generation are defined. The number of power steps depends on the level of accuracy that is required. Then a vector of all possible combinations of generator productions is produced. Among the combinations, those that are violating the UC constraints or are not within the hourly net demand range will be removed. The remaining combinations will be sorted based on the total values of the quadratic cost function of their respective generator outputs. Expensive combinations, that can safely be assumed that UC optimization problem will not elect, will also be removed. The obtained dataset only includes feasible and cheaper solutions. These solutions are expected to be around the optimal solution of the UC problem. In addition, considering solutions around the optimal one accounts for deviations from the planned generation schedule during real time operation. This dataset can be used as the training dataset for the intended ML methods.
\paragraph{Labelling the Data:}\label{sec:labelling}
This section briefly presents the SFR model used to analyze the frequency stability of small isolated power systems, which is used as a tool to label the features of the training dataset. The model is able to reflect the underlying short-term frequency response of small isolated power systems (like the La Plama Island system, under study). Note that any other dynamic power system model could be used. \Cref{fSFRmodel} details the power-system model typically used to design UFLS schemes for an island power system, consisting of $I$ generating units.
Each generating unit $i$ is represented by a second-order model approximation of its turbine-governor system. In fact, dynamic frequency responses are dominated by rotor and turbine-governor system dynamics. Excitation and generator transients can be neglected as they are much faster than the turbine-governor dynamics. The overall response of loads can be considered by means of a load-damping factor $D$ if its value is known. The gain $k_i$, which is inverse of the droop, and parameters $a_{i,1}$, $a_{i,2}$, $b_{i,1}$ and $b_{i,2}$, of each generating unit $i$ can be deduced from more accurate models or field tests. Gain $k_i$ is an important parameter in indicating the frequency response of unit $i$. Since it is important in the learning process of the ML model to have features that are able to represent the frequency dynamics after the outage, a weighted gain after the outage is defined here, which will later be used as a feature for the training dataset.
\begin{equation}
    \mathcal{K}_\ell=\sum\limits_{i\in \mathcal{I}, i\neq l}(k_i\mathcal{M}_i u_{t,i})
\end{equation}
Since primary spinning reserve is finite, power output limitations $\Delta p_{i,min}$ and $\Delta p_{i,max}$ are forced, so the units can only participate as much as their available reserve. Moreover, the ramp-up speed of the units should be limited to maximum ramping capacity of each respective unit. The complete model is explained in \cite{Sigrist2016}. 
\paragraph{Learning the Model:}
Considering the features $x \in \mathcal{X}$ and labels $y \in \{-1,+1\}$, with $+1$ for acceptable data points and $-1$ for unacceptable data points. 
The purpose of the ML model is to learn a decision function $f_{\theta}(x)$ which is positive when the label is $+1$ and negative when the label is $-1$, whilst minimizing misclassifications. Here, $\theta$ parametrizes the class of decision functions considered. For the purpose of this paper, the label indicates whether the resulting frequency nadir after an outage is tolerable or not. As the classifier is going to be implemented in the UC problem to be solved with MILP solvers, only decision functions of the following form are considered (with $\Theta\coloneqq(\theta_1,\dots,\theta_M)$):
\begin{equation}\label{eq:linearHS}
    f_{\theta}(x)\coloneqq
    \Theta^\intercal x+\theta_0
\end{equation}
Once $\theta_0$ and $\Theta$ have been trained from the data, \cref{eq:linearHS} can be directly added to the MILP formulation, simply by adding the follow constraint:
\begin{equation}\label{eq:MLconst}
    \Theta^\intercal x+\theta_0\geq 0
\end{equation}

\paragraph{Loss Function:}

The loss value $\ell(f_{\theta}(x),y)$ is the discrepancy between the true label $y$ and the sign of the decision function $f_{\theta}(x)$. The loss function measures how good the model predicts the actual outcome. 
We will find a classifier that minimizes the empirical risk (defined as the average loss value across the training data) plus a regularization term (if need be),
\begin{equation}
L_\theta\coloneqq
\frac{1}{N}\sum_{n=1}^N
\ell(f_\theta(x_n),y_n)
+
\frac{1}{C}\sum_{m=1}^M \theta_m^2
\end{equation}
where we assume the training data has $N$ samples.
$C$ represents a regularization parameter.
With smaller $C$ regularization is more effective, hence the model will be less prone to overfitting. With a larger $C$ the number of misclassifications on the training data might reduce, but at the cost of overfitting.

Different ML methods use different loss functions. For the purpose of this paper two ML methods suitable for binary classification are applied to the training dataset, namely logistic regression (LR) and support vector machines (SVM).
LR uses the log loss, without regularization ($C=\infty$):
\begin{equation}\label{eq:LR}
    \ell(f_{\theta}(x),y)\coloneqq
    -(1+y)\log\left(\frac{1}{1+e^{-f_{\theta}(x)}}\right)
    -(1-y)\log\left(\frac{e^{-f_\theta(x)}}{1+e^{-f_{\theta}(x)}}\right)
\end{equation}
The SVM model in this paper uses the regularized hinge loss function:
\begin{equation}
    \ell(f_{\theta}(x),y)\coloneqq
    \max\bigl(0,1-y f_{\theta}(x)\bigr)
\end{equation}

\paragraph{Validation:}
A standard way of validating the ML models is by cross validation. Cross validation is a statistical method of evaluating and comparing learning models by dividing the training dataset into two segments; one used to learn the model and the other used to validate the model. Cross validation is used to check the quality of ML models in \cref{results}.

\subsection{Evaluating the Methods}
\begin{figure}[!htbp]
\centering
\includegraphics[width=0.7\linewidth]{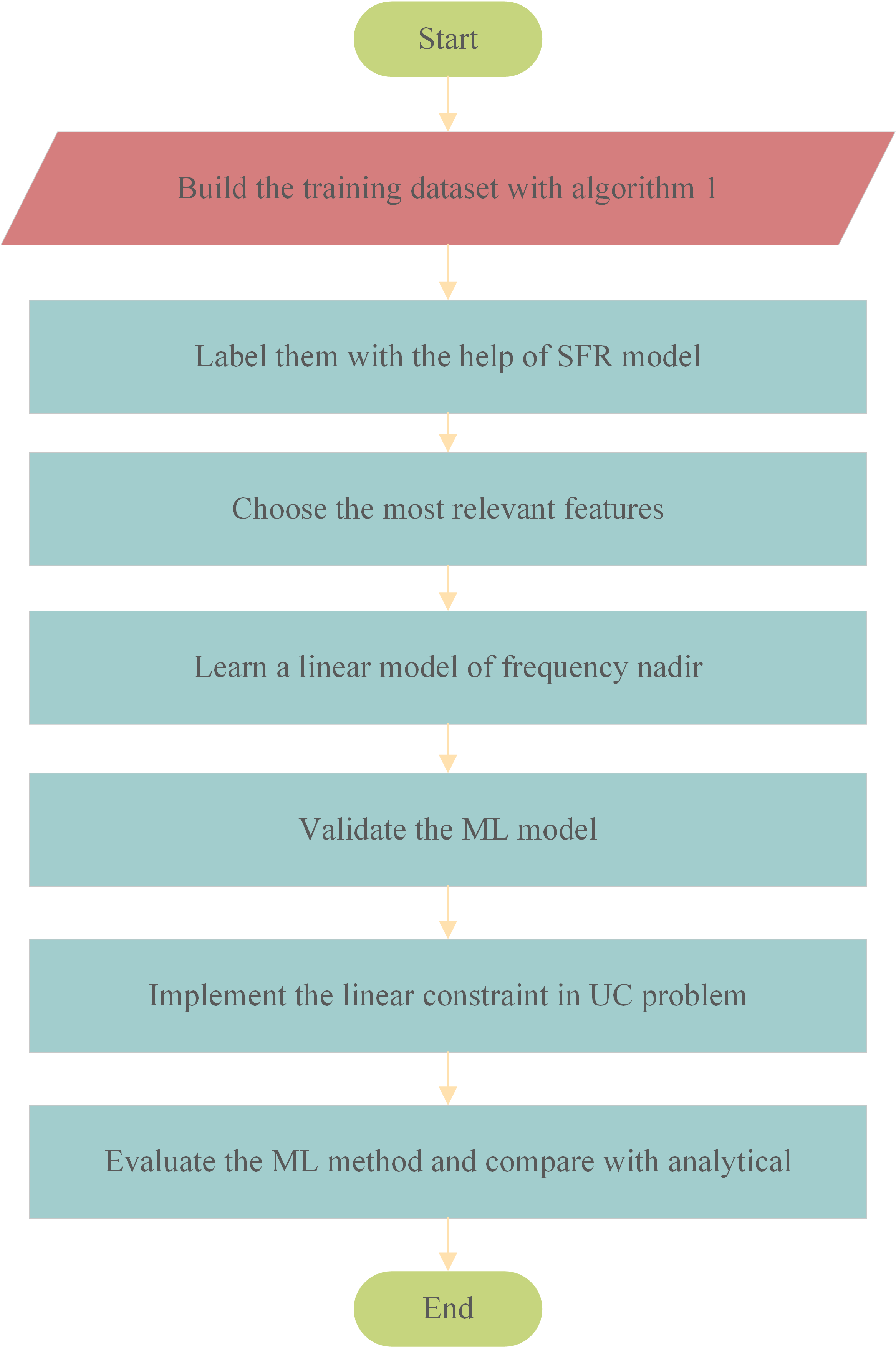}
\caption{Flowchart of the ML based methods}
\label{fig:flowchart}\end{figure}

A good method is able to ensure the frequency dynamics after outages, while keeping the operation cost low, and with a formulation that is computationally affordable. In the results the proposed ML method of including frequency nadir in UC is compared with analytical method and a base case (no frequency nadir constraint). To evaluate each of the methods, UC operation cost (as a representative of the costs), the amount of UFLS (as a representative of frequency dynamic quality), average frequency nadir after outage (as a representative of frequency dynamic quality), and the solution time of each method (as a representative of computational burden) are compared. A flowchart of the methodology is presented in \cref{fig:flowchart}.

\section{Results}\label{results}

\subsection{Case study}

\begin{figure}[!htbp]
\centering
\includegraphics[width=0.7\linewidth]{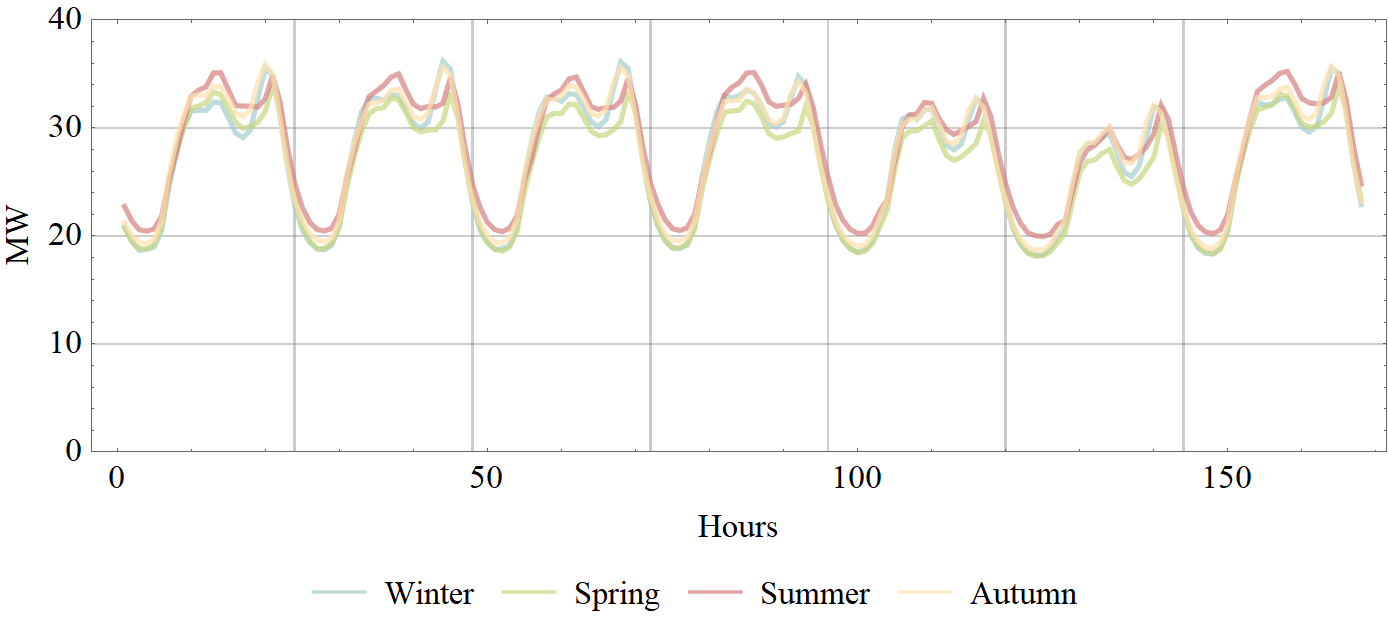}
\caption{Weekly demand for each season}
\label{fig:demand}\end{figure}

\begin{figure}[t]
\centering
\includegraphics[width=0.7\linewidth]{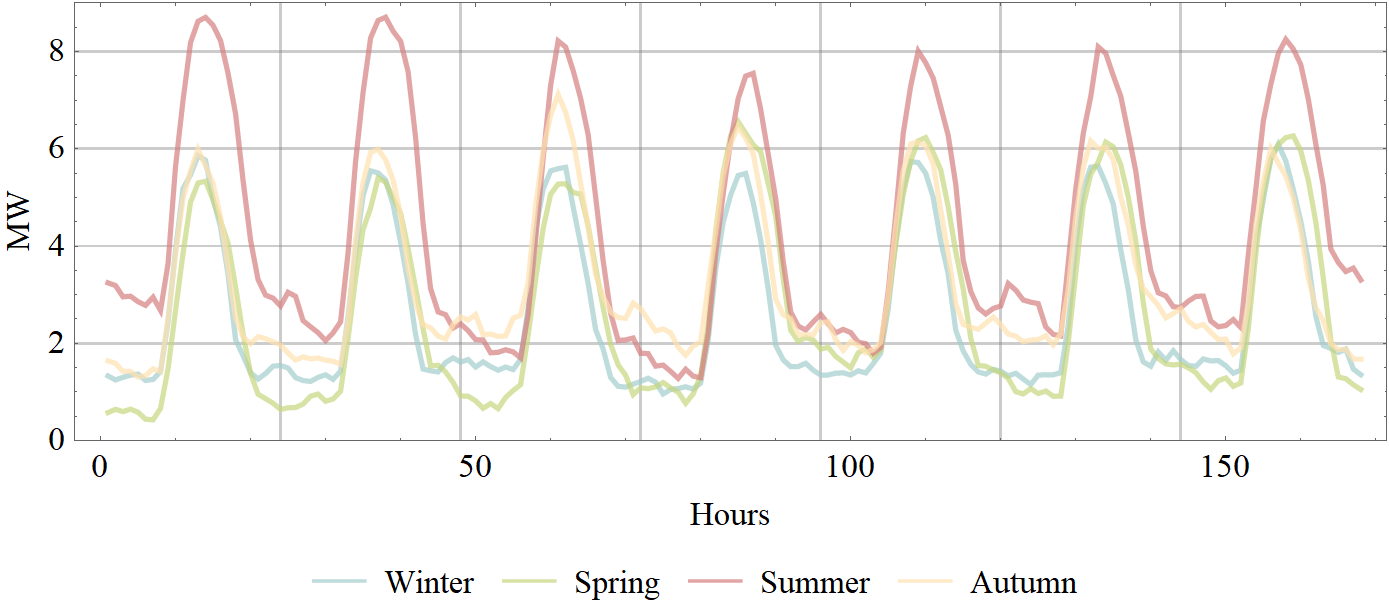}
\caption{Weekly available RES for each season}
\label{fig:res}\end{figure}

Simulations and comparisons of the methods are carried out on the real power system of the La Palma island, one of Spain’s Canary Islands. The yearly demand in 2018 is reported about 277.8 GWh
(average hourly demand of 31.7 MWh), supplied by eleven Diesel
generators pre-dominantly. According to \cite{de2019consejeria}, the installed capacity of the La Palma island power system mounts to 117.7 MW, where about 6\% of the installed capacity belongs to wind power generation. RES covers about 10\% of the yearly demand. The input data of the demand and availability of the RES to solve the UC problem is obtained from the most recent real data. Weekly demand of each sample week is shown in \cref{fig:demand}, and weekly available RES is shown in \cref{fig:res}.
To compare the methods, weekly UC is solved for sample weeks of different seasons. Doing so reveals any temporal dependency of the ML approaches to the training dataset. All the codes, input data, and results of this paper can be found at \url{https://doi.org/10.5281/zenodo.7082627}.

\subsection{Training Dataset}\label{sec:ResTraining}

\Cref{algori} is used to build a training dataset for La Palma island. Steps of 0.5 MW is used to define the vector of power levels. Then all possible combinations of operation points are generated. Among all the combinations, those that are either bigger than the annual thermal generation peak or smaller than the annual thermal generation minimum are excluded. Considering the historical data, thermal generation in La Palma island is between $\overline{\mathcal{D}}=36$ MW and $\underline{\mathcal{D}}=16$ MW, throughout the year. The training dataset should only include those operation points that are between maximum and minimum thermal generation. This training dataset is built to train the frequency nadir constraint. There is therefore no point to include any operation points that are violating other UC constraints, as they are not feasible. Notwithstanding this point, the operating points that are unable to provide enough reserve or cannot maintain minimum RoCoF constraint should be excluded as well. Amongst the remaining data points, those that are far from the optimal solution of the UC because of their incurred costs will never be selected as an optimal solution and therefore there is no point in keeping them either.

The remaining operating points are then sorted by the total value of their respective quadratic generation cost functions, and for every thermal generation level, those that are cheaper are kept. For the purpose of this paper, 500 operating points are kept for every thermal generation level. This final set of data points are considered as feasible data points and will be used as the training dataset for the frequency nadir constraint. All points should be labelled with the SFR model, as explained in \cref{sec:labelling}. To this end, all the data points are fed into the SFR model and frequency dynamics of every single generator outage will be obtained. The number of possible outages for the training dataset exceeds 90,000. The criterion here is to label any outages frequency nadir deviation more than 3.5 Hz as unacceptable (labeled with -1), and other outages as acceptable (labeled with 1). A summary of building and labelling the training dataset is presented in \cref{tab:datagen}. The process of building the training dataset can be updated annually.
\begin{table}[!htbp]
    \centering
    \begin{tabular}{c|c|c}\toprule
         &  & run time \\\hline
      generating data   & $19,500$ operation points  & $2,811''$ \\\hline
       labelling data  & $90,001$ single outages & $38,400''$\\
    \bottomrule
    \end{tabular}
    \caption{Summary of building and labeling the training dataset}\label{tab:datagen}
\end{table}

It is  important to define relevant features for the data points, that can represent the frequency nadir. Pearson correlation between frequency nadir and the chosen features for this study are shown in \cref{tab:correlation}. It is also stated that how each of the features represent for different operating points.
\begin{table}[!htbp]
    \centering
    \begin{tabular}{c|cccc}\toprule
      feature   & $x_1$ & $x_2$ & $x_3$ & $x_4$ \\\hline
      value   & $\mathcal{H}_\ell$ & $\mathcal{K}_\ell$ & $P_\ell$ & $\mathcal{R}_\ell$ \\\hline
      correlation  & $0.45$ & $0.47$  & $-0.81$ & 0.41 \\
    \bottomrule
    \end{tabular}
    \caption{Pearson correlation of chosen features and frequency nadir}\label{tab:correlation}
\end{table}
It's interesting to mention that weighted inertia after outage ($\mathcal{H}_\ell$) and weighted gain after outage ($\mathcal{K}_\ell$) are more correlated with frequency nadir, in comparison with available reserve ($\mathcal{R}_\ell$). Traditionally available reserve constraint has been the only criteria in UC problem to ensure the frequency stability after outages. Observations like what \cref{tab:correlation} shows, confirm that other than available reserve, more frequency dynamic related terms like available inertia and gain (which is inverse of the droop of the unit) should be taken into account too.

\subsection{Learning and Validating the Model}

\begin{figure}[!htbp]
\centering
\includegraphics[width=\linewidth]{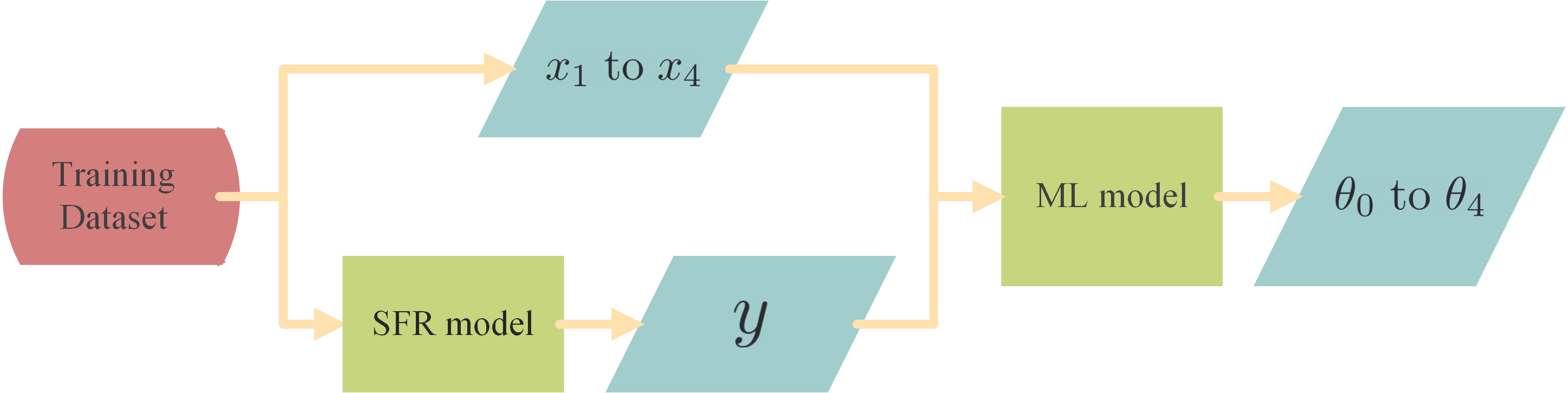}
\caption{The machine learning process.}
\label{fig:MLprocess}\end{figure}

Using different ML methods on the training dataset which consists of features in \cref{tab:correlation} and the labels from the SFR model, results in the learned decision function to be used in the UC problem. Different ML methods that are applied to the training dataset, their obtained decision function, and their corresponding training times and cross validations are summarized in \cref{tab:learning}. For cross validation purposes, the training dataset is randomly divided into a temporary test set, including 30\% of the whole data and the model is trained with the rest of the data. The percentage in the table is the accuracy of the model in predicting the labels of the test set.
\begin{table}[!htbp]
    \centering
    \begin{tabular}{c|ccccc|cc}\toprule
        & $\theta_0$ & $\theta_1$ & $\theta_2$ & $\theta_3$ & $\theta_4$ & training time & cross validation\\\hline
      LR & $1$ & $0.084$ & $ -0.013$ & $0.626$ & $-0.115$ & $0.4''$ & $96.7\%$ \\\hline
      SVM, C=1 & $1$ & $0.059$ & $-0.012$ & $ 0.806$ & $-0.154$ & $58.7''$ & $96.5\%$ \\\hline
      SVM, C=0.1 & $1$ & $0.062$ & $-0.012$ & $0.718$ & $-0.129$ & $92.6''$ & $96.6\%$ \\\hline
      SVM, C=10 & $1$ & $0.058$ & $-0.012$ & $  0.795$ & $ -0.152$ & $61.5''$ & $96.5\%$ \\
    \bottomrule
    \end{tabular}
    \caption{Learning process and results of ML methods.}\label{tab:learning}
\end{table}
As it can be seen in \cref{tab:learning}, training for the LR method is very fast. The SVM method can train the model in order of minutes. For LR and SVM, scikit-learn package in Python is used \cite{pedregosa11a}. The learning process is presented in \cref{fig:MLprocess}.

\subsection{Evaluating the Methods}

\begin{figure}[!b]
\centering
\includegraphics[width=0.8\linewidth]
{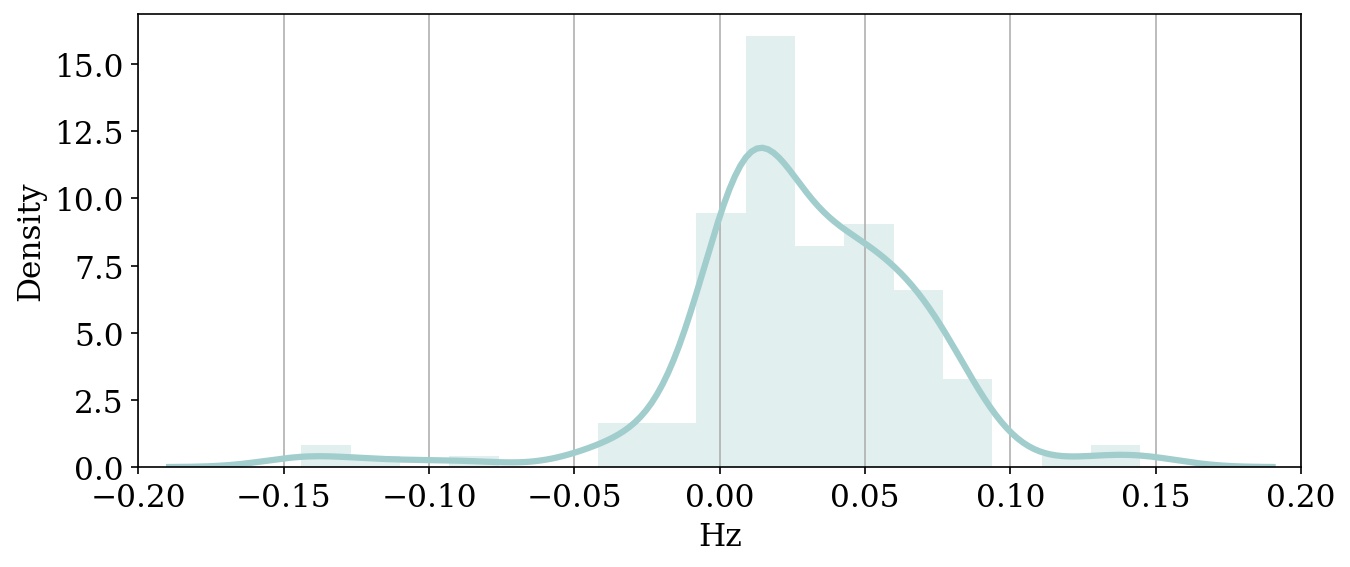}
\caption{Histogram of the difference between frequency nadir by eq. (9) and its approximation by separable programming for a sample week.}
\label{fig:comparisonNadir1}\end{figure}

\begin{figure}[!t]
\centering
\includegraphics[width=0.8\linewidth]{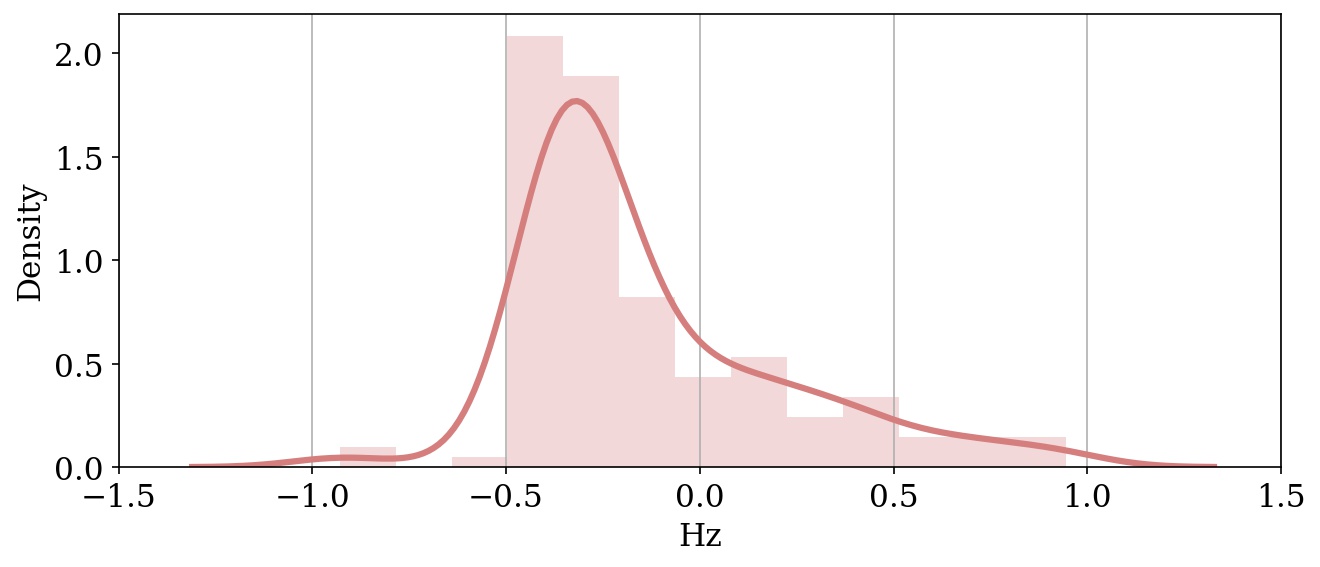}
\caption{Histogram of the difference between the frequency nadir from the SFR model and the frequency nadir approximation by separable programming for a sample week.}
\label{fig:comparisonNadir2}\end{figure}

In the analytical method, the frequency nadir for each outage is estimated by a close approximation of \cref{eq:nadir}. In \cref{fig:comparisonNadir1} the frequency nadir calculated by \cref{eq:nadir}, its approximation by separable programming approach are compared, for all the outages in a sample day.
This figure confirms that the linearized approximation of frequency nadir is very accurate. The maximum error of approximation is only $0.15$ Hz. In \cref{fig:comparisonNadir2} the difference between the frequency nadir from the SFR model and the frequency nadir approximation by separable programming is shown on a histogram for a sample week.
As shown in \cref{fig:comparisonNadir2}, the analytical method is underestimating the frequency drop, compared to what is obtained from the SFR model. The main reason for this is the underlying assumption in \cref{eq:reslin}. \Cref{eq:reslin} assumes that all the units will deliver their available reserve linearly in $T_g$ seconds, regardless of their actual response speed.

All ML methods of learning frequency nadir that are presented in \cref{tab:learning} and the analytical method, which is explained in \cref{sec:fNadir}, are applied to the UC problem. For the ML methods the set of constraints that are defined in \cref{eq:MLconst} are added as the frequency constraint. However, for the analytical method, all the equations that are defined in \cref{sec:fNadir} to approximate the frequency nadir equation, should be added to the UC problem. Weekly UC is solved for sample weeks of winter, spring, summer and autumn. Input demand and available RES are obtained from historical data and are shown in figures \ref{fig:demand} and \ref{fig:res}. In \cref{tab:ucResults} all the methods are compared.
\begin{table}[!htbp]
    \centering
    \begin{tabular}{c|ccccc}\toprule
        & operation cost(k\euro) & UFLS/outage (MW) & $f^{nadir}/outage (Hz)$ & run-time \\\hline
      base case & $824.08$ & $1.682$ & $-1.298$ 
      & $317''$  \\\hline
      LR & $829.38$ & $1.182$ & $-1.146$ 
      & $302''$  \\\hline
      SVM, C=1 & $829.41$ & $1.130$ & $-1.127$ 
      & $295''$  \\\hline
      SVM, C=0.1 & $829.72$ & $1.182$ & $-1.140$ 
      & $261''$  \\\hline
      SVM, C=10 & $829.52$ & $1.230$ & $-1.151$ 
      & $277''$ \\\hline
      analytical & $829.60$ & $1.355$ & $-1.167$ & $25,506''$\\
    \bottomrule
    \end{tabular}
    \caption{Average weekly UC results for different methods.}\label{tab:ucResults}
\end{table}
Some indicators are presented in \cref{tab:ucResults} to compare the performance of each method. The ultimate purpose is to minimize the operation cost and the UFLS, with a formulation that is computationally affordable. The operation cost in \cref{tab:ucResults}, indicates the weekly expenditure on electricity generation. UFLS/outage indicates the average amount of load shedding after single outages. This is a good indicator for the quality of frequency response after the outage. Methods that have smaller load shedding per outage, are better able to prevent severe outages.

\begin{figure}
    \centering
    \includegraphics[width=0.7\linewidth]{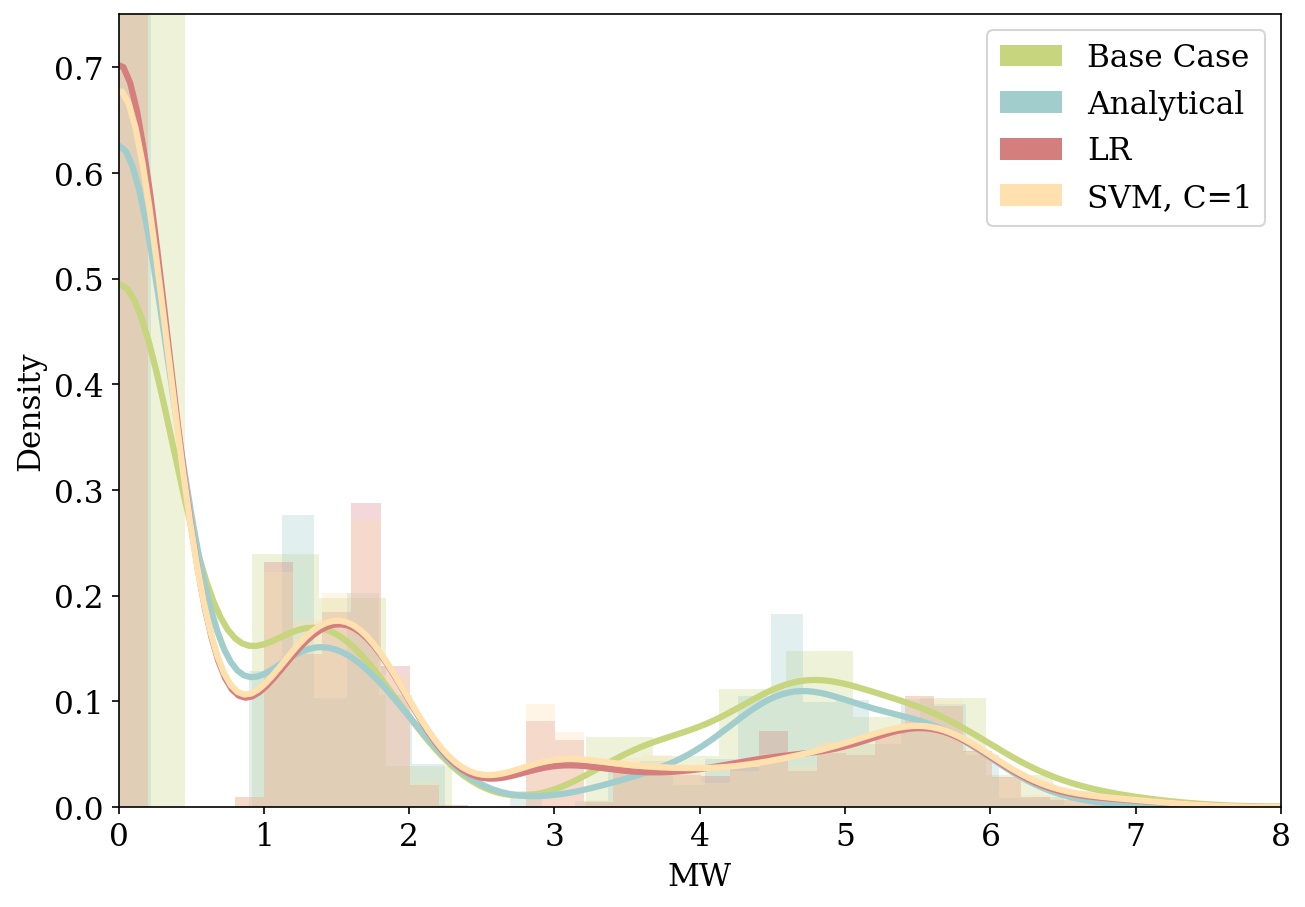}
    \caption{Histogram of UFLS for different methods.}
    \label{fig:ufls_histo}
\end{figure}

A histogram of the UFLS activation for different methods is presented in \cref{fig:ufls_histo}.
\Cref{fig:ufls_histo} shows that the LR and SVM methods, have been able to prevent the activation of UFLS more than the others. Also, they've considerably decreased the number of incidents with big amount of UFLS activation.
The average run-time of weekly UC is presented in \cref{tab:ucResults} as an indicator of computational efficiency of the methods. Furthermore, the average of frequency nadir after outages is presented in the table. This shows how each method manages to restrict the severity of frequency nadir. There is a considerable difference in the run-time of ML based methods with the analytical method. The studied ML base methods in this paper, only introduce one constraint for the outage of each generator and each time interval to represent frequency nadir constraint in UC. But to represent frequency nadir analytically, an approximation of \cref{eq:nadir} suitable for MILP must be formulated. To handle the non-convexity of \cref{eq:nadir}, a full set of constraints and very many additional auxiliary binary variables (see \cref{eq:linearization}) must be introduced, adding substantial computational complexity. As the UC problem is usually solved weekly and daily for short-term scheduling, it's important to keep the solution time as low as possible. The results in \cref{tab:ucResults} confirm that the ML based methods are much faster than the analytical method.

Among the ML based methods, LR has led to the lowest operation cost, while keeping the load shedding low. The SVM method with $C=10$ has the highest amount of load shedding per outage, among the ML based methods. As expected, the SVM with big $C$ tends to overfit. Another interesting point to mention is that although the ML based methods are prone to misclassifying some of the incidents, the amount of average load shedding is lower than the analytical approach. The reason is that the analytical method's prediction of frequency nadir, is based on the assumption in \cref{eq:reslin} and then it is approximated by a piecewise linear function. It is therefore seen that both analytical and ML based methods have their own sources of error. On the other hand, analytical methods are independent of any training dataset. This is a big advantage, because the training dataset depends on the inputs of UC problem and the topology of the system. Once the topology of the system changes (e.g. a new generator is installed) or the inputs change radically (e.g. the annual demand or availability of RES changes), the training dataset and the ML model should be updated. Another downside of ML methods is the lack of trust in the ML methods, due to their black-box nature. A summary is presented in \cref{tab:MLVsAna} that compares ML based methods with analytical ones.\par
\begin{table}[!htbp]
    \centering
    \begin{tabular}{p{1.7cm}|p{4.5cm}p{4.5cm}}\toprule
        & \centering ML based & \centering analytical \tabularnewline\hline
      advantages &-it can be solved in a timely manner\newline-it doesn't increase the size of the problem too much \newline-it can be used in more complicated formulations of UC, like robust and stochastic models  & -it's independent of training data\newline-it's directly obtained from the physics of frequency dynamics in power systems\\\hline
      disadvantages & -it depends on training dataset which should be updated once in a while\newline-labelling the dataset can be hard\newline-operators might be sceptical about ML based methods & -it imposes a lot of new constraint and variables to UC problem\newline-UC solution time is high\newline-implementing it in robust and stochastic models of UC is challenging\\\hline
      source of error & -misclassification of the model when it's applied to real inputs\newline -inaccuracies in labeling\newline-ill-defined dataset   & -fixed time reserve response assumption\newline-approximate due to piecewise linearization  \\
    \bottomrule
    \end{tabular}
    \caption{Comparison of ML based and analytical methods}\label{tab:MLVsAna}
\end{table}

\section{Conclusion}\label{conclusion}
This paper uses LR and SVM to classify outages with tolerable and intolerable frequency nadir. We compared this against a piecewise linear approximation of the frequency nadir that uses separable programming. Both approaches were then tested on the same test system (La Palma Island in Spain) for purposes of solving frequency-constrained UC problems as mixed-integer linear programming problems. The piecewise linearized formulation of frequency nadir is computationally much more demanding. The results of the comparison study show that our ML based methods are as accurate as the piecewise linear formulation, without the added computational burden, in preventing the outages that exceed critical frequency nadir. This is important in building confidence further in using ML based methods in safety critical applications in power systems such as solving FCUC problems. Meanwhile solving UC with ML based frequency nadir constraint is considerably faster, results in much less computational expenditure, and allows for more flexible assumptions on system response. 

\bibliographystyle{IEEEtran}
\bibliography{bib}\label{references}
\end{document}